\documentclass[12pt,draftclsnofoot,onecolumn]{IEEEtran}
\usepackage[cmex10]{amsmath}
\usepackage{amssymb}
\usepackage{graphicx}
\newtheorem{theorem}{Theorem}
\newtheorem{proposition}{Proposition}
\newtheorem{lemma}{Lemma}
\newcommand{\dotleq}{\buildrel \textstyle .\over \leq}

\begin{document}
\title{\LARGE On the Achievable Diversity-Multiplexing Tradeoff in MIMO Fading Channels With Imperfect CSIT}
\author{Xiao Juan~Zhang and Yi~Gong
\thanks{The authors are with the School
of Electrical and Electronic Engineering, Nanyang Technological
University, Singapore, 639798 (e-mail: zh0012an@ntu.edu.sg, eygong@ntu.edu.sg).}}
\maketitle
\begin{abstract}
In this paper, we analyze the fundamental tradeoff of diversity and
multiplexing in multi-input multi-output (MIMO) channels with
imperfect channel state information at the transmitter (CSIT). We
show that with imperfect CSIT, a higher diversity gain as well as a
more efficient diversity-multiplexing tradeoff (DMT) can be
achieved. In the case of multi-input single-output
(MISO)/single-input multi-output (SIMO) channels with $K$
transmit/receive antennas, one can achieve a diversity gain of
$d(r)=K(1-r+K\alpha)$ at spatial multiplexing gain $r$, where
$\alpha$ is the \emph{CSIT quality} defined in this paper. For
general MIMO channels with $M$ ($M>1$) transmit and $N$ ($N>1$)
receive antennas, we show that depending on the value of $\alpha$,
different DMT can be derived and the value of $\alpha$ has a great
impact on the achievable diversity, especially at high multiplexing
gains. Specifically, when $\alpha$ is above a certain threshold, one
can achieve a diversity gain of $d(r)=MN(1+MN\alpha)-(M+N-1)r$;
otherwise, the achievable DMT is much lower and can be described as
a collection of discontinuous line segments depending on $M$, $N$,
$r$ and $\alpha$. Our analysis reveals that imperfect CSIT
significantly improves the achievable diversity gain while enjoying
high spatial multiplexing gains.
\end{abstract}

\newpage
\section{Introduction}

The performance of wireless communications is severely degraded by
channel fading caused by multipath propagation and interference from
other users. Fortunately, multiple antennas can be used to increase
diversity to combat channel fading. Antenna diversity where
sufficiently separated or different polarized multiple antennas are
put at either the receiver, the transmitter, or both, has been
widely considered \cite{Paulraj}, \cite{Lek}. On the other hand,
multi-antenna channel fading can be beneficial since it can increase
the degrees of freedom of the channel and thus can provide spatial
multiplexing gain. 
It is shown in \cite{FoschiniGans} that the spatial multiplexing
gain in a multi-input and multi-output (MIMO) Rayleigh fading
channel with $M$ transmit and $N$ receive antennas increases
linearly with $\min(M,N)$ if the channel knowledge is known at the
receiver. As MIMO channels are able to provide much higher spectral
efficiency and diversity gain than conventional single-antenna
channels, many MIMO schemes have been proposed, which can be
classified into two major categories: spatial multiplexing oriented
(e.g., Layered space-time architecture \cite{Foschini}), and
diversity oriented (e.g., space-time trellis coding \cite{Tarokh},
\cite{Yi}, and space-time block coding \cite{Alamouti},
\cite{Orthogonal}).

For a MIMO scheme realized by a family of codes \{$C(\rho)$\} with
signal-to-noise ratio (SNR) $\rho$, rate $R(\rho)$ (bits per channel
use), and maximum-likelihood (ML) error probability
$\mathcal{P}_e(\rho)$, Zheng and Tse defined in \cite{IEEErelay:DMT}
the spatial multiplexing gain $r$ as
$r\triangleq\lim_{\rho\rightarrow\infty} \frac{R(\rho)}{\log \rho}$
and the diversity gain $d$ as
$d\triangleq-\lim_{\rho\rightarrow\infty}
\frac{\mathcal{P}_e(\rho)}{\log \rho}$. Under the assumption of
independent and identically distributed (i.i.d.) quasi-static flat
Rayleigh fading channels where the channel state information (CSI)
is known at the receiver but not at the transmitter, for any integer
$r\leq \min(M,N)$, the optimal diversity gain $d^*(r)$ (the supremum
of the diversity gain over all coding schemes) is given by
\cite{IEEErelay:DMT}
\begin{equation}
\label{DMT} d^*(r)=(M-r)(N-r)
\end{equation}
provided that the code length $L\geq M+N-1$. The
diversity-multiplexing tradeoff (DMT) in (\ref{DMT}) provides a
theoretical framework to analyze many existing diversity-oriented
and multiplexing-oriented MIMO schemes. It indicates that the
diversity gain cannot be increased without penalizing the spatial
multiplexing gain and vice versa. This pioneering work has generated
a lot of research activities in finding DMT for other important
channel models \cite{IEEErelay:MAC}-\cite{IEEErelay:FDTCR} and
designing space-time codes that achieve the desired tradeoff of
diversity and multiplexing gain
\cite{IEEErelay:ODASC}-\cite{IEEErelay:RDTST}. The DMT analysis was
extended to multiple-access channels in \cite{IEEErelay:MAC}. The
automatic retransmission request (ARQ) scheme is shown to be able to
significantly increase the diversity gain by allowing
retransmissions with the aid of decision feedback and power control
in block-fading channels \cite{IEEErelay:ARQ}. The work in
\cite{IEEErelay:FDPRM} investigated the diversity performance of
rate-adaptive MIMO channels at finite SNRs and showed that the
achievable diversity gains at realistic SNRs are significantly lower
than those at asymptotically high SNRs. The impact of spatial
correlation on the DMT at finite SNRs was further studied in
\cite{IEEErelay:FDTCR}.

It is natural to expect that the DMT can be further enhanced through
power and/or rate adaptation if the transmitter has channel
knowledge. If the CSI at the transmitter (CSIT) is perfectly known,
there will be no outage even in slow fading channels since it is
always able to adjust its power or rate adaptively according to the
instantaneous channel conditions. For example, it can transmit with
a higher power or lower rate when the channel is poor and a lower
power or higher rate when the channel is good. However, in practice
the CSIT is almost always imperfect due to imperfect CSI feedback
from the receiver or imperfect channel estimation at the transmitter
through pilots. The work in \cite{IEEErelay:TCTT} showed that the
transmitter training through pilots significantly increases the
achievable diversity gain in a single-input
 multi-output (SIMO) link. In \cite{IEEErelay:OFT}, the authors quantified
the CSIT quality as $\alpha=-\log \sigma_{e}^2/\log \rho$, where
$\sigma_{e}^2$ is the variance of the CSIT error, and showed that
using rate adaptation, one can achieve an average diversity gain of
$\bar{d}(\alpha,\bar{r})=(1+\alpha-\bar{r})K$ in SIMO/MISO links,
where $K=\max(M,N)$ and $\bar{r}$ is the average multiplexing gain.
Note that setting $\alpha=1$ and ignoring the multiplexing gain loss
due to training symbols directly yields the result in
\cite{IEEErelay:TCTT}. For general MIMO channels, the achievable DMT
with partial CSIT  is characterized in \cite{IEEErelay:TT}, where
the partial CSIT is obtained using quantized channel feedback.

In this paper, we analyze the fundamental DMT in MIMO channels and
show that with power adaptation, imperfect CSIT provides significant
additional diversity gain over (\ref{DMT}). The imperfect CSIT
considered in this paper is due to channel estimation error at the
transmitter side. In the case of MISO/SIMO channels, we show that
with power adaptation (under an average sum power constraint), one
can achieve a higher diversity gain than that with rate adaptation
in \cite{IEEErelay:OFT}, where the authors assumed peak power
transmission and thus no temporal power adaptation is considered
therein. Specifically, we prove that with a CSIT quality $\alpha$,
the achievable diversity gain is $d(r)=K\left(1-r+K\alpha\right)$.
It has been shown in our earlier work \cite{SIMOMISO} that this is
actually the \emph{optimal} DMT in SIMO/MISO channels with CSIT
quality $\alpha$. For general MIMO channels ($M>1$, $N>1$), we show
in this paper that depending on the value of $\alpha$, different DMT
can be derived and the value of $\alpha$ has a great impact on the
achievable diversity, especially at high multiplexing gains.
Specifically, when $\alpha$ is above a certain threshold, one can
achieve a diversity gain of $d(r)=MN(1+MN\alpha)-(M+N-1)r$;
otherwise, the achievable DMT is much lower and can be described as
a collection of discontinuous line segments depending on $M$, $N$,
$r$ and $\alpha$. It is noted that an independent and concurrent
work recently reported in \cite{NoisyCSIT} shares some similar
results. However, we wish to
emphasize that 
our CSIT model and the involved analysis towards the achievable DMT
are different from those in \cite{NoisyCSIT}. The noisy CSIT therein
is based on the \emph{channel mean feedback} model in
\cite{MeanFeedback} and an example of obtaining CSIT through delayed
feedback is provided, whereas the CSIT in our work is estimated from
reverse channel pilots using ML estimation at the transmitter. As
the variance of the channel estimation error is inversely
proportional to the pilots' SNR \cite{IEEErelay:SDICE}, the CSIT
quality $\alpha$ is naturally connected to the reverse channel power
consumption and any value of $\alpha$ can be achieved by scaling the
reverse channel transmit power.
In addition, our
paper provides detailed closed-form solutions to the achievable DMT,
which offers great insight and depicts directly what the DMT curve
with imperfect CSIT looks like.



\emph{Notations}: $\mathcal{R}^N$ denotes the the set of real
$N$-tuples, and $\mathcal{R}^{N+}$ denotes the set of non-negative
$N$-tuples. Likewise, $\mathcal{C}^{N\times M}$ denotes the set of
complex $N\times M$ matrices. For a real number $x$, $(x)^+$ denotes
$\max(x,0)$, while for a set $\mathcal{O}\subseteq \mathcal{R}^N$,
$\mathcal{O}^+$ denotes $\mathcal{O}\cap\mathcal{R}^{N+}$.
${\mathcal{O}}^c$ denotes the complementary set of $\mathcal{O}$ and
$\emptyset$ denotes the empty set. $|\mathcal{O}|$ denotes the
cardinality of set $\mathcal{O}$. $x\in (a, b]$ denotes that the
scalar $x$ belongs to the interval $a < x \leq b$. Likewise, $x\in
[a, b]$ is similarly defined. $\mathcal{CN}(0,\sigma^2)$ denotes the
complex Gaussian distribution with mean 0 and variance $\sigma^2$.
The superscripts $^*$ and $^{\dag}$ denote the complex conjugate and
conjugate transpose, respectively. $\|\cdot\|_F^2$ denotes the
matrix Frobenius norm and $\textbf{I}_N$ denotes the $N\times N$
identity matrix. $E\{\cdot\}$ denotes the expectation operator and
$\log(\cdot)$ denotes the base-2 logarithm. $f(\rho)\doteq \rho^b$
denotes that $b$ is the exponential order of $f(\rho)$, i.e.,
$\lim_{\rho\rightarrow\infty} {\log(f(\rho))}/{\log(\rho)}=b$.
Likewise, $\dotleq$ is similarly defined. Finally, for matrix
$\mathbf{A}$, $\mathbf{A} \succeq 0$ denotes that $\mathbf{A}$ is
positive semidefinite; if $\succeq$ is used with a vector, it
denotes the componentwise inequality.

The rest of this paper is organized as follows. In section II,  we
describe the channel model. In section III, we propose a power
adaptation scheme based on imperfect CSIT and present the main
result on the achievable DMT. The achievability proof of the
presented DMT is given in Section IV. Section V provides some
discussions. Finally, Section VI concludes this paper.

\section{Channel Model}

We consider a point-to-point TDD wireless link with $M$ transmit and
$N$ receive antennas, where the downlink and uplink channels are
reciprocal. Without loss of generality, we assume $M\geq N$ in this
paper. As shown in \cite{IEEErelay:DMT}, this assumption does not
affect the DMT result. We also consider quasi-static Rayleigh fading
channels, where the channel gains are constant within one
transmission block of $L$ symbols, but change independently from one
block to another. We assume that the channel gains are independently
complex circular symmetric Gaussian with zero mean and unit
variance. The channel model, within one block, can be written as
\begin{equation}
\label{SIMO_channel}
\textbf{Y}=\sqrt{{P}/M}\textbf{H}\textbf{X}+\textbf{W}
\end{equation}
where $\textbf{H}=\{h_{n,m}\}\in\mathcal{C}^{N\times M }$ with
$h_{n,m}$, $m=1,2,\ldots,M$, $n=1,2,\ldots,N$, being the channel
gain from the $m$-th transmit antenna to the $n$-th receive antenna;
$\textbf{X}=\{X_{m,l}\} \in \mathcal{C}^{M\times L}$ with $X_{m,l}$,
$m=1,2,\ldots,M$, $l=1,2,\ldots,L$, being the symbol transmitted
from antenna $m$ at time $l$; $\textbf{Y}=\{Y_{n,l}\}\in
\mathcal{C}^{N\times L}$ with $Y_{n,l}$, $n=1,2,\ldots,N$,
$l=1,2,\ldots,L$, being the received signal at antenna $n$ and time
$l$; the additive noise $\textbf{W}\in \mathcal{C}^{N\times L}$ has
i.i.d. entries $W_{n,l}\sim \mathcal{CN}(0,\sigma^2)$; $P$ is the
instantaneous transmit power while the average energy of $X_{m,l}$
is normalized to be 1. Letting $\bar{P}$ denote the average sum
power constraint, we have $E\{P\}=\bar{P}$. So, the average SNR at
the receive antenna is given by $\rho=\bar{P}/\sigma^2$.

We assume that the receiver has perfect CSI $\textbf{H}\in
\mathcal{C}^{N\times M}$, but the transmitter has imperfect CSIT
$\hat{\textbf{H}}\in \mathcal{C}^{N\times M}$, which is estimated
from reverse channel pilots using ML estimation. Thus,
$\hat{\textbf{H}}$ can be modeled as 
\cite{IEEErelay:SDICE}-\cite{IEEErelay:SDVA}
\begin{equation}
\hat{\textbf{H}}=\textbf{H}+\textbf{E}
\end{equation}
where the channel estimation error $\textbf{E}\in
\mathcal{C}^{N\times M}$ has i.i.d. entries $E_{n,m}\sim
\mathcal{CN}(0,\sigma_{e}^2)$, $n=1,2,\ldots,N$, $m=1,2,\ldots,M$,
and is independent of $\textbf{H}$. The quality of
$\hat{\textbf{H}}$ is thus characterized by $\sigma_{e}^2$. If
$\sigma_{e}^2 = 0$, the transmitter has perfect channel knowledge;
if $\sigma_{e}^2$ increases, the transmitter has less reliable
channel knowledge. We follow \cite{IEEErelay:OFT} to quantify the
channel quality at the transmitter. The transmitter is said to have
a \emph{CSIT quality} $\alpha$, if $\sigma_{e}^2\doteq
\rho^{-\alpha}$. The definition of $\alpha$ builds up a connection
between the imperfect channel knowledge at transmitters and the
forward channel SNR, $\rho$. Since the variance of the channel
estimation error is inversely proportional to the pilots' SNR, i.e.,
$\sigma_{e}^2\propto (SNR_{pilot})^{-1}$ \cite{IEEErelay:SDICE}, any
value of $\alpha$ can be achieved by scaling the reverse channel
power such that $SNR_{pilot}\doteq\rho^{\alpha}$. One can see that
the selection of
 $\alpha$ value actually determines the cost of obtaining
 CSIT in terms of the reverse channel power consumption. When $\alpha=0$,
 the reverse channel SNR does not
 scale with $\rho$, which means that the pilot power is fixed or limited; when $0<\alpha<1$,
 the reverse channel SNR relative
 to $\rho$ is asymptotically zero; when $\alpha=1$, the
 reverse channel SNR scales with $\rho$ at the same rate; when $\alpha>1$,
 the reverse channel SNR as compared to the forward channel SNR, $\rho$, is asymptotically unbounded \cite{IEEErelay:OFT}.
In the sequel, we will study how the pilot power, or equivalently
the CSIT quality $\alpha$, affects the fundamental tradeoff of
diversity and multiplexing in the considered channel. Before
presenting our main results, we give the following probability
density function (pdf) expressions and some preliminary results that
will be used later.

For an $N\times M$ ($N\leq M$) random matrix $\textbf{A}$ with
i.i.d. entries $\sim \mathcal{CN}(0,1)$, let $0<\lambda_1 \leq
\lambda_2 \leq ... \leq \lambda_N$
 denote the ordered nonzero eigenvalues of $\textbf{A}\textbf{A}^{\dag}$. Letting $v_n$ denote the exponential order of
$1/\lambda_n$ for all $n$, the pdf of the random vector
$\textbf{v}=[v_1,...,v_N]$ is given by \cite{IEEErelay:Eigen}
\begin{equation}
\label{eqnpv2}
\begin{split}
p(\textbf{v})&=\lim_{\rho\rightarrow\infty}\xi^{-1}(\log \rho)^N\prod_{n=1}^N \rho^{-(M-N+1)v_n} \prod_{j>n}^N (\rho^{-v_n}-\rho^{-v_j})^2\exp\left(-\sum_{n=1}^N \rho^{-v_n}\right)\\
&\doteq
\begin{cases}
0,\ \text{for any $v_n<0$}\\
\prod_{n=1}^N \rho^{-(2n-1+M-N)v_n},\  \text{for all $v_n\geq0$}
\end{cases}
\end{split}
\end{equation}where $\xi$ is a normalizing constant. Hence, the probability $\mathcal{P}_\mathcal{O}$ that
$(v_1,...,v_N)$ belongs to set $\mathcal{O}$ can be characterized by
\begin{equation}
\label{bb} \mathcal{P}_\mathcal{O}\doteq \rho^{-d_{\mathcal{O}}},\
\text{for}\ d_{\mathcal{O}}= \inf_{(v_1,...,v_N)\in \mathcal{O}^+}
\sum_{n=1}^N (2n-1+M-N)v_n
\end{equation}provided that $\mathcal{O}^+$ is not empty.

Letting $\textbf{a}=[a_1,a_2...,a_N]$, $0<a_1\leq a_2\leq...\leq
a_N$, $\textbf{b}=[b_1,b_2...,b_N]$, $0<b_1\leq b_2\leq...\leq b_N$,
and $\textbf{c}=[c_1,c_2...,c_N]$, $0<c_1\leq c_2\leq...\leq c_N$,
denote the eigenvalue vectors of $\textbf{H}\textbf{H}^{\dag}$,
$\hat{\textbf{H}}\hat{\textbf{H}}^{\dag}$ and $
\textbf{E}\textbf{E}^{\dag}$, respectively, the pdfs of
$\textbf{a}$, $\textbf{b}$, and $\textbf{c}$ can be shown to be
\begin{eqnarray}
\label{eqn_9}
&&p(\textbf{a})=\xi^{-1}\prod_{n=1}^N a_n^{M-N} \prod_{n<j}^N (a_n-a_j)^2\exp\left(-\sum_{n=1}^N a_n\right)\\
&&p(\textbf{b})=\hat{\xi}^{-1}
\prod_{n=1}^N b_n^{M-N} \prod_{n<j}^N (b_n-b_j)^2\exp\left(-\frac{1}{1+\sigma_e^2}\sum_{n=1}^N b_n\right) \\
\label{eqn_11} &&p(\textbf{c})=\tilde{\xi}^{-1} \prod_{n=1}^N
c_n^{M-N} \prod_{n<j}^N
(c_n-c_j)^2\exp\left(-\frac{1}{\sigma_e^2}\sum_{n=1}^N c_n\right)
\end{eqnarray}
where $\hat{\xi}^{-1}=\xi^{-1} (1+\sigma_e^2)^{-MN}$ and
$\tilde{\xi}^{-1}=\xi^{-1} (\sigma_e^2)^{-MN}$.

\section{Main Result on DMT}
The ML error probability $\mathcal{P}_e(\rho)$ of the channel
described in (\ref{SIMO_channel}) is closely related to the
associated packet outage probability $\mathcal{P}_{out}$, which is
defined as the probability that the instantaneous channel capacity
falls below the target data rate $R(\rho)$. In fact, the error
probability of an ML decoder which utilizes a fraction of the
codeword such that the mutual information between the received and
transmitted signals exceeds $LR(\rho)$ (no outage), averaged over
the ensemble of random Gaussian codes, can be made arbitrarily small
provided that the codeword length $L$ is sufficiently large
\cite{IEEErelay:OAD}. We will thus leverage on the outage
probability to examine the achievable diversity gain. If the
transmitter has perfect CSIT, it may adopt the optimal power
adaptation according to the actual instantaneous channel gain such
that no outage will occur. With only the imperfect CSIT, in order to
mitigate the channel uncertainty, we propose the following power
adaptation scheme.

\begin{proposition}
Given $\hat{\textbf{H}}$, the transmitter
 transmits with power
 \begin{equation}
 \label{PA}
P(\hat{\textbf{H}})=\frac{ \kappa\bar{P}}{\left(\prod_{n=1}^N
b_n^{2n-1+M-N}\right)^t}
\end{equation}where $\kappa=\hat{\xi}\prod_{n=1}^N \left[(2n-1+M-N)(1-t)\right]$ and
 $t$ ($0\leq t<1$) can be chosen arbitrarily close to $1$.
\end{proposition}

It is shown in Appendix A that the above power adaptation scheme
satisfies the sum power constraint $E\{P(\hat{\textbf{H}})\}=
\bar{P}$. We believe that given the CSIT quality of $\alpha$, this
power adaptation scheme is the optimal power adaptation scheme that
maximizes the achievable diversity gain of a MIMO fading channel.

\begin{theorem}

Consider a MIMO channel with $M$ transmit and $N$ receive antennas
($M\geq N$) and CSIT quality of $\alpha$. If the block length $L\geq
M+N-1$, the achievable DMT using the power adaptation scheme in
Proposition 1 is characterized by

Case 1: If $N=1$ or $\alpha \geq \frac{1}{M-1}$, then
\begin{equation}
\begin{split}
d(r)=MN(1+MN\alpha)-(M+N-1)r.
\end{split}
\label{eq14}
\end{equation}

Case 2: Otherwise, the achievable DMT is a collection of
discontinuous line segments, with the two end points of line segment
$d_k(r)$ ($k\in \mathcal{B}$) given by
\begin{equation}
\label{eqn_446}
\begin{split}
\text{Left end: }&d_k(r)=k(M-N+k){\tau}(k),\  \text{for}\ r=(N-k){\tau}(k)\\
\text{Right end: }&d_k(r)=((N-k)(k-N-1)+MN){\tau}(k)-(2k-1+M-N)(N-{\mathcal{I}}(k)){\tau}({\mathcal{I}}(k)),\\
&\ \ \ \ \ \ \ \ \ \ \text{for}\
r=(N-{\mathcal{I}}(k)){\tau}({\mathcal{I}}(k))
\end{split}
\end{equation}where
\begin{equation}
\mathcal{B}=\left\{k\middle | (M-N+k)(N-k)<1/\alpha,
(N-k){\tau}(k)<(N-\bar{k}){\tau}(\bar{k}), \forall \bar{k}<k,
k=1,...,N\right\}, \nonumber
\end{equation}
${\tau}(k)=1+k\alpha(M-N+k)$ and ${\mathcal{I}}(k)= \max_{\bar{k}\in
{\mathcal{B}},\bar{k}<k} \bar{k}$. 
\end{theorem}

For example, when $M=N=2$ and $\alpha<1$, the DMT curve consists of
two discontinuous line segments which are
$(0,16\alpha+4)$---$(1+\alpha,13\alpha+1)$ and
$(1+\alpha,1+\alpha)$---$(2,2\alpha)$. When $r=1+\alpha$, the
achievable diversity gain is $d(r)=1+\alpha$ instead of
$13\alpha+1$. From Theorem 1, we can get $d(0)=MN(1+MN\alpha)$ and
$d(N)=p\alpha(M-N+p)(MN+(p-N)(N-p+1)))-p^2+p$ where $p=\min_{k\in
{\mathcal{B}}} k$. If $\alpha <\frac{1}{(N-1)(M-N+1)}$, which
indicates $1\in {\mathcal{B}}$, we will have $d(N)=\alpha
N(M-N+1)^2$.

\section{Proof of Theorem 1}
The proof involves the computation of the asymptotic ML  error
probability at high SNRs. We will first derive a lower bound of the
SNR exponent of the outage probability, denoted as
$d_\mathcal{O}(r)$, and then show that using a random coding
argument the SNR exponent of the error probability is no less than
$d_\mathcal{O}(r)$ if $L\geq M+N-1$.

\subsection{Derivation of $d_{\mathcal{O}}(r)$}
Optimizing over all input distributions, which  can be assumed to be
Gaussian with a covariance matrix $\textbf{Q}$ without loss of
optimality, the outage probability of a MIMO channel with transmit
power $P(\hat{\textbf{H}})$ is given by
\begin{equation}
\label{OutageProb} \mathcal{P}_{out}=\inf_{\textbf{Q}\succeq0,
tr(\textbf{Q})\leq M}\mathcal{P}\left(\log
\text{det}\left(\textbf{I}_N+\frac{P(\hat{\textbf{H}})}{M\sigma^2}\textbf{H}\textbf{Q}\textbf{H}^{\dag}\right)<R(\rho)
\right)
\end{equation}
where $\mathcal{P}(\cdot)$ denotes the probability of an event. It
is shown in \cite{IEEErelay:DMT} that one can get an upper bound and
a lower bound on the outage probability by picking
$\textbf{Q}=\textbf{I}_{M}$ and $\textbf{Q}=M\textbf{I}_M$,
respectively, and the two bounds converge in the high SNR regime.
Therefore, without loss of generality, we consider
$\textbf{Q}=\textbf{I}_M$. Substituting (\ref{PA}) in
(\ref{OutageProb}), we have
\begin{equation}
\begin{split}
\mathcal{P}_{out}&=\mathcal{P}\left(\log \text{det}\left(\textbf{I}_N+\frac{\rho \kappa}{M\prod_{n=1}^N b_n^{(2n-1+M-N)t}}\textbf{H}\textbf{H}^{\dag}\right)<R(\rho) \right)\\
&=\mathcal{P}\left(\log\prod_{n=1}^N\left(1+\frac{\rho \kappa
a_n}{M\prod_{n=1}^N b_n^{(2n-1+M-N)t}}\right)<R(\rho) \right).
\end{split}
\end{equation}

\begin{lemma}The eigenvalues of $\hat{\textbf{H}}\hat{\textbf{H}}^{\dag}$,
$\textbf{H}\textbf{H}^{\dag}$ and $\textbf{E} \textbf{E}^{\dag}$
have the following relationship
\begin{equation}
\label{bac} b_n \leq 2 (a_n+c_N), \  n=1,2,...,N.
\end{equation}
\end{lemma}
\begin{IEEEproof}
We obviously have the following equality
\begin{equation}
(\textbf{H}+\textbf{E})(\textbf{H}+\textbf{E})^{\dag}+(\textbf{H}-\textbf{E})(\textbf{H}-\textbf{E})^{\dag}=2 (\textbf{H}\textbf{H}^{\dag}+\textbf{E}\textbf{E}^{\dag})
\end{equation}where both $(\textbf{H}+\textbf{E})(\textbf{H}+\textbf{E})^{\dag}$
and $(\textbf{H}-\textbf{E})(\textbf{H}-\textbf{E})^{\dag}$ are
positive semidefinite matrices. We denote the vector of eigenvalues
of $(\textbf{H}\textbf{H}^{\dag}+\textbf{E}\textbf{E}^{\dag})$ as
$\textbf{d}=[d_1,...,d_N]$ with $d_1\leq d_2\leq...\leq d_N$. Since
the eigenvalues of the sum of two positive semidefinite matrices are
at least as large as the eigenvalues of any one of the positive
semidefinite matrices \cite{IEEErelay:Math}, we have $ b_n\leq 2
d_n,\  n=1,2,...,N$. Further, using the relationship of the
eigenvalues of the sum of Hermitian matrices, we get $a_n + c_1 \leq
d_n \leq a_n +c_N,\  n=1,2,...,N$. It thus directly leads to
(\ref{bac}).
 \end{IEEEproof}

With Lemma 1, the outage probability is upper bounded by
\begin{equation}
\label{out} \mathcal{P}_{out}\leq
 \mathcal{P}\left[\log\prod_{n=1}^N \left(1+ \frac{\rho \kappa
a_n}{M\prod_{n=1}^N (2a_n+2c_N)^{(2n-1+M-N)t}}
\right)<R(\rho)\right] .
\end{equation}

Let $v_n$ and $u_n$ denote the exponential orders of $1/{a_n}$ and
$1/{c_n}$, respectively, i.e.,
$v_n=-\lim_{\rho\rightarrow\infty}\frac{\log(a_n)}{\log(\rho)}$,
$u_n=-\lim_{\rho\rightarrow\infty}\frac{\log(c_n)}{\log(\rho)}$.
Using (\ref{eqnpv2}), (\ref{eqn_9}) and (\ref{eqn_11}), the pdfs of
the random vector $\textbf{v}=[v_1,...,v_N]$ and
$\textbf{u}=[u_1,...,u_N]$ can be shown to be
\begin{eqnarray}
&&p(\textbf{v})\doteq
\begin{cases}
0,\ \text{for any $v_n<0$}\\
\prod_{n=1}^N \rho^{-(2n-1+M-N)v_n},\  \text{for all $v_n\geq0$}
\end{cases}\label{pvn1}\\
&&p(\textbf{u})\doteq
\begin{cases}
0,\ \text{for any $u_n<\alpha$}\\
\prod_{n=1}^N \rho^{-(2n-1+M-N)(u_n-\alpha)},\  \text{for all
$u_n\geq \alpha$}.
\end{cases}\label{pun1}
\end{eqnarray}

At high SNRs, with (\ref{pvn1}) and (\ref{pun1}), (\ref{out})
becomes
\begin{equation}
\label{eqn29} \mathcal{P}_{out}\leq \mathcal{P}\left[\sum_{n=1}^N
\left(1-v_n+\sum_{n=1}^N t(2n-1+M-N)\min(v_n,u_N)\right)^+
<r\right].
\end{equation}

So, the outage event $\mathcal{O}$ in (\ref{eqn29}) is the set of
$\{v_1,\ldots,v_N,u_1,\ldots,u_N\}$ that satisfies
\begin{equation}
\label{eqn28} \sum_{n=1}^N \left(1-v_n+\sum_{n=1}^N
t(2n-1+M-N)\min(v_n,u_N)\right)^+ <r
\end{equation}
where $ v_n\geq 0, u_n\geq \alpha\geq0, n=1,2,...,N$.

According to (\ref{bb}), we have
\begin{equation}
\mathcal{P}_{out}\leq \mathcal{P}_\mathcal{O}\doteq
\rho^{-d_{\mathcal{O}}(r)}
\end{equation}
where $d_\mathcal{O}(r)$ serves as a lower bound of the SNR exponent
of $\mathcal{P}_{out}$ and is given by
\begin{equation}
\label{DMTNEW} d_\mathcal{O}(r)= \inf_{(v_1,...,v_N,u_1,...,u_N)\in
\mathcal{O}} \sum_{n=1}^N (2n-1+M-N) \left(v_n+u_n-\alpha\right).
\end{equation}

Next, we work on the explicit expression of $d_\mathcal{O}(r)$.
Since the left hand side (LHS) of (\ref{eqn28}) is a non-decreasing
function of $u_N$, decreasing $u_N$ will not violate the outage
condition in (\ref{eqn28}) while enjoying a reduced SNR exponent
$\sum_{n=1}^N (2n-1+M-N) \left(v_n+u_n-\alpha\right)$. Combining
with the fact $u_n\geq \alpha, \ n=1,2,...,N$, the solution of
$\textbf{u}$ is found to be $u_1^*=...=u_N^*=\alpha$. Therefore,
(\ref{eqn28}) can be rewritten as
 \begin{equation}
 \mathcal{O}=\left\{v_n\middle|\sum_{n=1}^N \left(1-v_n+\sum_{n=1}^N t(2n-1+M-N)\min(v_n,\alpha)\right)^+ <r, v_n\geq 0 \right\}.
 \end{equation}

To solve the optimization problem of (\ref{DMTNEW}), we need to solve the subproblems
\begin{equation}
\label{eqn33} d_k(r)\triangleq \inf_{(v_1,...,v_N)\in \mathcal{O}_k}
\sum_{n=1}^N (2n-1+M-N) v_n,\ k=0,1,...,N
\end{equation}
where subset $\mathcal{O}_k$ $(0\leq k \leq N)$ is defined as
\begin{equation}
\begin{split}
\mathcal{O}_k  =&\left\{v_n\middle|\sum_{n=1}^N
\left(1-v_n+\sum_{n=1}^k t(2n-1+M-N)\alpha + \sum_{n=k+1}^N
t(2n-1+M-N)v_n\right)^+ <r,\right.\nonumber\\ &\left.v_1\geq...\geq
v_k\geq \alpha \geq v_{k+1}\geq...\geq v_N \right\} .
 \end{split}
\end{equation}

So, $d_\mathcal{O}(r)$ is given by
\begin{equation}
\label{eqn_29} d_\mathcal{O}(r)=\min\left( d_0(r),
d_1(r),...,d_N(r)\right).
\end{equation}
In other words, among all the DMT curves $d_0(r)$,...,$d_N(r)$,
corresponding to the outage subsets
$\mathcal{O}_1$,...,$\mathcal{O}_N$, the lowest one will be the DMT
curve for the entire outage event. Since $t$ can be made arbitrarily
close to 1, it is without loss of accuracy to set $t=1$ in the rest
of this paper.

Firstly, we derive $d_0(r)$. 
It is easy to show
 \begin{equation}
 \label{eqn_30}
\sum_{n=1}^N \left(1-v_n+\sum_{n=1}^N (2n-1+M-N)v_n\right)^+\geq
N-\sum_{n=1}^N v_n +N\sum_{n=1}^N (2n-1+M-N)v_n \geq N
\end{equation}
which suggests that it is possible to operate at spatial
multiplexing gain $r\in[0,N]$ reliably without any outage, i.e.,
$d_0(r)=\infty$. So we can exclude $d_0(r)$ from the optimization
problem in (\ref{eqn_29}).

Secondly, we derive $d_k(r)$ ($1\leq k \leq N$). Note that the function
 $\sum_{n=1}^N \left(1-v_n+\sum_{n=1}^k (2n-1\right.$ $\left.+M-N)\alpha + \sum_{n=k+1}^N (2n-1+M-N)v_n\right)^+$
  is an increasing function of $v_{k+1},v_{k+2},...,v_N$. That is,
  decreasing $v_{k+1},v_{k+2}...,v_N$ does not violate the outage condition for
  $\mathcal{O}_k$, while reducing the SNR exponent $\sum_{n=1}^N (2n-1+M-N)v_n$.
  Therefore, the optimal solutions of $v_{k+1},v_{k+2},...,v_N$ are
  $v_{k+1}^*=v_{k+2}^*=...=v_N^*=0$. Consequently, the
  optimization problem in (\ref{eqn33}) can be reformulated as
 \begin{equation}
 d_k(r)=\inf_{(v_1,...,v_k)\in \tilde{\mathcal{O}}_k} \sum_{n=1}^k (2n-1+M-N) v_n,\
 k=1,...,N.
 \end{equation}
Here the modified outage subset $\tilde{\mathcal{O}}_k$ is defined
as
\begin{equation}
\label{eqn35}
\begin{split}
\tilde{\mathcal{O}}_k
=\left\{v_1,...,v_k\middle|N\tau(k)-\sum_{n=1}^k v_n<r, \alpha\leq
v_k\leq...\leq v_1 \leq \tau(k) \right\}.
\end{split}
\end{equation}
where $\tau(k)=1+k\alpha(M-N+k)$. Careful observation of
(\ref{eqn35}) reveals that
\begin{equation}
\label{rr}
\begin{split}
 N\tau(k)-\sum_{n=1}^k v_n \geq (N-k)\tau(k).
\end{split}
\end{equation}
It implies that there will be no outage ($d_k(r)=\infty$), if $r
\leq (N-k)\tau(k)$ or $(N-k)\tau(k) \geq N$. Note that $(N-k)\tau(k)
\geq N \Rightarrow (M-N+k)(N-k)\geq 1/\alpha$. So, if $(M-N+k)(N-k)<
1/\alpha$, there will be nonzero outage ($d_k(r)<\infty$), for $r\in
\Omega_k$, where $\Omega_k$ is defined as
\begin{equation}
\label{ome} \Omega_k\triangleq \left((N-k)\tau(k),\ N\right].
\end{equation}

For any $r\in \Omega_k$, we are able to explicitly calculate the
optimal solutions of $v_1,...,v_k$ that minimize the SNR exponent
$\sum_{n=1}^k (2n-1+M-N)v_n$. The results are summarized in the
following.

\begin{enumerate}
    \item If $r=(N-k')\tau(k)-(k-k')\alpha,\
k'=1,2,...,k$, then the achievable diversity for outage event
$\tilde{\mathcal{O}}_k$ is
\begin{equation}
\label{eqn_35} d_k(r)=k'(M-N+k')\tau(k)+(k-k')(k+k'+M-N)\alpha.
\end{equation} The corresponding optimal solutions of $v_1,...v_k$ are
$v_1^*=...=v_{k'}^*=\tau(k)$, $v_{k'+1}^*=...=v_k^*=\alpha$.
Specifically, $d_k(r)=k(M-N+k)\tau(k)$ for $r=(N-k)\tau(k)$.

\item  If
$(N-k')\tau(k)-(k-k')\alpha<r<(N-k'+1)\tau(k)-(k-k'+1)\alpha,\
k'=1,2,...,k$,  the achievable diversity for outage event
$\tilde{\mathcal{O}}_k$ is
\begin{equation}
\label{eqn_38}
d_k(r)=((N-k')(k'-N-1)+MN)\tau(k)+(k-k'+1)(k-k')\alpha-(2k'-1+M-N)r.
\end{equation}
The corresponding optimal solutions of $v_1,...,v_k$ are
$v_1^*=...=v_{{k'-1}}=\tau(k)$,
$v_{k'}^*=(N-k'+1)\tau(k)-(k-k')\alpha-r$,
$v_{k'+1}^*=...=v_k^*=\alpha$.
 \end{enumerate}

%
For a particular $k'$, when spatial multiplexing gain $r$ is
 between $(N-k')\tau(k)-(k-k')\alpha$ and
 $(N-k'+1)\tau(k)-(k-k'+1)\alpha$, only one singular value
 of $\textbf{H}$, corresponding to the typical outage event, needs to be adjusted
 to be barely large enough to support the data rate. (\ref{eqn_38}) further shows that
 $d_k(r)$ is a continuous line segment between these two points. It is thus obvious that curve $d_k(r)$ is piecewise-linear with
$(r,d_k(r))$ specified in (\ref{eqn_35}) being its corner points.

After calculating $d_1(r),...,d_N(r)$, we remain to solve
$d_{\mathcal{O}}(r)=\min_k d_k(r),\ k=1,...,N$. Since
$d_k(r)=\infty$ if $(M-N+k)(N-k)\geq 1/\alpha $, we only need to
consider $k\in \mathcal{A}$, where set $\mathcal{A}$ is defined as
\begin{equation}
\mathcal{A}=\left\{k\middle | (M-N+k)(N-k)< 1/\alpha,
k=1,...,N\right\}.
\end{equation}

Note that we always have $k=N \in \mathcal{A}$. We consider the
following two cases.

Case 1: $\mathcal{A}$ has only one element, i.e.,
$\mathcal{A}=\{k=N\}$. In this case, we have
$d_{\mathcal{O}}(r)=d_N(r)$. If $N=1$, this condition is naturally
satisfied, since there is only one element in $\mathcal{A}$ that is
$k=1$. If $N>1$, we must require $(M-N+k)(N-k) \geq 1/\alpha$ for
$k=1,...,N-1$, which leads to
\begin{equation}
\label{eqn42} \alpha  \geq \frac{1}{M-1},\ \ N>1.
\end{equation}

We now examine the corner points of $d_N(r)$. From (\ref{eqn_35}),
we have $r=(N-k')(1+\alpha MN-\alpha)> 1+MN\alpha-\alpha$ for corner
point $k'$ ($k'=1,2,...,N-1$). Since $1+MN\alpha-\alpha$ is a
non-decreasing function of $\alpha$, we easily get $r>
1+\frac{MN}{M-1}-\frac{1}{M-1} >N$. Thus we conclude that there is
only one corner point ($0,d_N(0)$) on curve $d_N(r)$ over region
$r\in \Omega_{N}$. Therefore, $d_{\mathcal{O}}(r)=d_N(r)$ is a
straight line between corner points ($0,d_N(0)$) and ($N,d_N(N)$).
From (\ref{eqn_38}), we have $d_N(N)=MN(1+MN\alpha)-(M+N-1)N$, so
$d_{\mathcal{O}}(r)$ can be described as
\begin{equation}
d_{\mathcal{O}}(r)=MN(1+MN\alpha)-(M+N-1)r \ \ \text{for}\ 0\leq
r\leq N.
\end{equation}

Case 2: $\mathcal{A}$ has more than one element. Since $N\in
\mathcal{A}$ and $\Omega_N=[0, N]$, $\Omega_k$ ($k\neq N,k\in
\mathcal{A}$) overlaps with $\Omega_N$. That is, there are some
regions of spatial multiplexing gain $r$, leading to finite
diversity gains on different DMT curves. A straightforward method to
find $d_{\mathcal{O}}(r)$ is to numerically calculate $d_k(r)$ for
all $k\in \mathcal{A}$, and choose the minimum value among them.
However, this makes $d_{\mathcal{O}}(r)$ implicit and hardly
insightful. To find the closed-form solution of
$d_{\mathcal{O}}(r)$, we wish to find out if there is any
relationship among $d_1(r),...,d_N(r)$. This motivates the birth of
the following Lemma, the proof of which is given in Appendix B.

\begin{lemma}
For any spatial multiplexing gain $r\in \Omega_{k_1} \bigcap \Omega_{k_2}$ ($1\leq k_1, k_2\leq N$), if $k_1<k_2$, we have $d_{k_1}(r)<d_{k_2}(r)$.
\end{lemma}

This Lemma tells us if a spatial multiplexing gain $r$ leads to
finite diversity gains on two DMT curves, we only need to select the
curve with lower diversity gain. For example, if $r\in
\Omega_1\bigcap\Omega_2\bigcap...\bigcap\Omega_N$, then
$d_{\mathcal{O}}(r)=d_1(r)$ since $d_1(r)<d_2(r)<...<d_N(r)<\infty$.
Therefore, we can further expurgate bad $k$ (s.t. $\Omega_k\subseteq
\Omega_{\bar{k}}$, for $\bar{k}<k \in \mathcal{A} $) from
$\mathcal{A}$ and only take into account $k\in {\mathcal{B}}$ for
the optimization problem, where
\begin{equation}
{\mathcal{B}}=\left\{k\middle |
(N-k)\tau(k)<(N-\bar{k})\tau(\bar{k}),\  \forall \bar{k}<k,
\bar{k},k \in \mathcal{A}\right\}.
\end{equation}

Letting $|{\mathcal{B}}|$ denote the cardinality of ${\mathcal{B}}$,
we further divide $r\in[0,N]$ into $|{\mathcal{B}}|$ non-overlapping
regions with region $\tilde{\Omega}_k$ ($k\in {\mathcal{B}}$)
defined as
\begin{equation}
\begin{split}
\tilde{\Omega}_k&=\Omega_k\bigcap \tilde{\Omega}_{\bar{k}}^c,\  \forall \bar{k}<k\&\bar{k}\in {\mathcal{B}}\\
&=
[(N-k)\tau(k), (N-\mathcal{I}(k))\tau(\mathcal{I}(k)))\\
\end{split}
\end{equation}
where $\mathcal{I}(k)$ indicates the immediately preceding element
of $k$ in ${\mathcal{B}}$, i.e.,
$\mathcal{I}(k)=\max_{\bar{k}<k,\bar{k}\in {\mathcal{B}}} \bar{k}$.
From Fig. 1, which illustrates the relationship between $\Omega_{k}$
and $\tilde{\Omega}_k$, we get $d_{\mathcal{O}}(r)=d_k(r)$ for any $
r\in \tilde{\Omega}_k$.

Next we examine the corner points on curve $d_k(r)$ over $r\in
\tilde{\Omega}_k$ and give the following Lemma, the proof of which
is given in Appendix C.
\begin{lemma}
For $k\in {\mathcal{B}}$, there is only one corner point, $\left(
(N-k)\tau(k), k(M-N+k)\tau(k)\right)$, making $r\in
\tilde{\Omega}_k$.
\end{lemma}

As a result, $d_k(r)$ over $r\in \tilde{\Omega}_k$ is just a single
line segment connecting the following two end points
\begin{equation}
\label{eqn_44}
\begin{split}
\text{Left end: }&d_k(r)=k(M-N+k)\tau(k),\  \text{for}\ r=(N-k)\tau(k)\\
\text{Right end: }&d_k(r)=((N-k)(k-N-1)+MN)\tau(k)-(2k-1+M-N)(N-\mathcal{I}(k))\tau(\mathcal{I}(k)),\\
&\ \ \ \ \ \ \ \ \ \text{for}\
r=(N-\mathcal{I}(k))\tau(\mathcal{I}(k)).
\end{split}
\end{equation}

Finally, since $d_{\mathcal{O}}(r)$ is the union of $d_k(r)$ over
$r\in \tilde{\Omega}_k$ for all $k\in {\mathcal{B}}$,
 the DMT curve over the entire outage event is the collection of all the involved
line segments and the two end points of line segment $d_k(r)$ ($k\in
{\mathcal{B}}$) are described in (\ref{eqn_44}). It should be noted
that these line segments are discontinuous though $r$ is continuous
between $0$ and $N$. Combining the above Cases 1 and 2 directly
leads to (\ref{eq14}) and (\ref{eqn_446}) in Theorem 1.


\subsection{Achievability Proof}

To complete the proof of the Theorem 1, we need to show that
$\mathcal{P}_e(\rho)\dotleq \rho^{-d_{\mathcal{O}}(r)}$ if $L\geq
M+N-1$. With the ensemble of i.i.d. complex Gaussian random codes at
the input, the ML error probability is given by \cite{IEEErelay:DMT}
\begin{equation}
\label{eqn41}
\mathcal{P}_e(\rho)=\mathcal{P}_{\mathcal{O}}\mathcal{P}(\text{error}
|\mathcal{O})+\mathcal{P}(\text{error},\mathcal{O}^c)\leq
\mathcal{P}_{\mathcal{O}}+\mathcal{P}(\text{error},\mathcal{O}^c)
\end{equation}where $\mathcal{O}$ and $\mathcal{P}_{\mathcal{O}}$ are given by (\ref{eqn28}) and (\ref{DMTNEW}), respectively.

$\mathcal{P}(\text{error},\mathcal{{O}}^c)$ can be upper-bounded by
a union bound. Assume that $\textbf{X}(0)$, $\textbf{X}(1)$ are two
possible transmitted codewords, and that $\Delta
\textbf{X}=\textbf{X}(1)-\textbf{X}(0)$. Suppose $\textbf{X}(0)$ is
transmitted, the probability that an ML receiver will make a
detection error in favor of $\textbf{X}(1)$, conditioned on a
certain realization of the channel, is
\begin{equation}
\mathcal{P}\left(\textbf{X}(0)\rightarrow \textbf{X}(1)\middle |
\textbf{H},\hat{\textbf{H}}\right)=\mathcal{P}\left(\frac{P(\hat{\textbf{H}})}{M
\sigma^2}\left\| \frac{1}{2}\textbf{H}(\Delta \textbf{X})\right
\|_F^2\leq \|\textbf{w}\|^2\right)
\end{equation}where $\textbf{w}$ is the additive noise on the direction of $\textbf{H}(\Delta \textbf{X})$, with variance $1/2$. With the standard approximation of the Gaussian tail function, $Q(x)\leq 1/2 \exp(-x^2/2)$, we have
\begin{equation}
\mathcal{P}\left(\textbf{X}(0)\rightarrow \textbf{X}(1)\middle |
\textbf{H},\hat{\textbf{H}}\right)\leq \exp
\left(-\frac{P(\hat{\textbf{H}})}{4M \sigma^2}\|\textbf{H}(\Delta
\textbf{X})\|^2\right).
\end{equation}

Averaging over the ensemble of random codes, we have the average
pairwise error probability conditioned on the channel realization
\begin{equation}
\mathcal{P}\left(\textbf{X}(0)\rightarrow \textbf{X}(1)\middle |
\textbf{H},\hat{\textbf{H}}\right)\leq \det
\left(\textbf{I}_N+\frac{P(\hat{\textbf{H}})}{2M
\sigma^2}\textbf{H}\textbf{H}^{\dag}\right)^{-L}.
\end{equation}

With a data rate $R=r\log(\rho)$ (bits per channel use), we have in total $\rho^{Lr}$ codewords. Applying the union bound, we have
\begin{equation}
\begin{split}
\mathcal{P}(\text{error} | \textbf{H},\hat{\textbf{H}})&\leq\rho^{Lr} \det \left(\textbf{I}_N+\frac{P(\hat{\textbf{H}})}{2M \sigma^2}\textbf{H}\textbf{H}^{\dag}\right)^{-L}\\
&= \rho^{Lr} \prod_{n=1}^N \left(1+ \frac{\rho \kappa
a_n}{2M\prod_{n=1}^N b_n^{2n-1+M-N}} \right)^{-L}
\\
&\leq \rho^{Lr}  \prod_{n=1}^N \left(1+ \frac{\rho \kappa
a_n}{M\prod_{n=1}^N (2a_n+2c_N)^{2n-1+M-N}}
\right)^{-L}\\
&\doteq \rho^{-L\left(\sum_{n=1}^N \left(1-v_n+\sum_{n=1}^N
(2n-1+M-N)\min(v_n,u_N)\right)^+-r\right)}.
\end{split}
\end{equation}

Averaging over the distributions of $\textbf{H}$ and $ \hat{\textbf{H}}$, or equivalently
 $\textbf{v}$ and $\textbf{u}$, we have
\begin{equation}
\begin{split}
&\mathcal{P}(\text{error}, \mathcal{{O}}^c)=\int_{\mathcal{O}^c}p(\textbf{u})p(\textbf{v}) \mathcal{P}(\text{error} | \textbf{H},\hat{\textbf{H}})d\textbf{u} d\textbf{v}\\
&\dotleq \int_{\mathcal{{O}}^c}\rho^{-\sum_{n=1}^N(2n-1+M-N)(v_n+u_n-\alpha)} \rho^{-L\left(\sum_{n=1}^N \left(1-v_n+\sum_{n=1}^N (2n-1+M-N)\min(v_n,u_N)\right)^+-r\right)} d\textbf{u} d\textbf{v}\\
&\doteq \rho^{-d_G(r)}
\end{split}
\end{equation}
where
\begin{equation}
\label{eqn_48}
\begin{split}
d_G(r)
=&\inf_{\textbf{u}, \textbf{v} \in \mathcal{O}^c}\sum_{n=1}^N(2n-1+M-N)(v_n+u_n-\alpha)\\
&+L\left(\sum_{n=1}^N \left(1-v_n+\sum_{n=1}^N
(2n-1+M-N)\min(v_n,u_N)\right)^+-r\right).
\end{split}
\end{equation}

When $L\geq M+N-1$, $d_G(r)$ has the same monotonicity as
$\sum_{n=1}^N \left(1-v_n+\sum_{n=1}^k (2n-1+M\right.$
$\left.-N)\alpha + \sum_{n=k+1}^N (2n-1+M-N)v_n\right)^+$ with
respect to $v_n$ or $u_n$, $n=1,...,N$. Therefore, the minimum
always occurs when $\sum_{n=1}^N \left(1-v_n+\sum_{n=1}^N
(2n-1+M-N)\min(v_n,u_N)\right)^+=r$. Hence
\begin{equation}
\begin{split}
d_G(r)&=\inf_{\sum_{n=1}^N \left(1-v_n+\sum_{n=1}^N (2n-1+M-N)\min(v_n,u_N)\right)^+=r} \sum_{n=1}^N(2n-1+M-N)(v_n+u_n-\alpha)\\
&=d_{\mathcal{O}}(r).
\end{split}
\end{equation}

Therefore, the overall error probability can be written as
\begin{equation}
\begin{split}
\mathcal{P}_e(\rho)&\leq \mathcal{P}_{\mathcal{O}}+\mathcal{P}(\text{error},\mathcal{{O}}^c)\\
&\doteq \rho^{-d_{\mathcal{O}}(r)}+\mathcal{P}(\text{error},\mathcal{{O}}^c)\\
& \dotleq \rho^{-d_{\mathcal{O}}(r)}+\rho^{-d_G(r)} \doteq
\rho^{-d_{\mathcal{O}}(r)}
\end{split}
\end{equation}

Since the MIMO channel with the proposed power adaptation scheme
leads to an error probability lower than or equal to
$\rho^{-d_{\mathcal{O}}(r)}$, we can say that the MIMO channel is
able to achieve the diversity gain of $d_{\mathcal{O}}(r)$. Theorem
1 is thus obtained.

\section{Discussions}

In this section, we discuss the additional diversity gain $\Delta_{
d}(r)$ brought by the imperfect CSIT through power adaptation.

Case A) $N=1$ (MISO/SIMO): According to (\ref{eq14}), the imperfect
CSIT provides an additional diversity gain of
$\Delta_{d}(r)=M^2\alpha $ at any spatial multiplexing gain in the
considered MISO/SIMO channel. Most remarkably, when $\alpha = 1/M$,
one can achieve both full diversity gain
 (i.e., $M$) and full spatial multiplexing gain (i.e., $1$) at the same time,
while $\alpha$ has to be equal to or greater than $1$ to achieve the
same performance in \cite{IEEErelay:OFT}. Note that the case of
$\alpha<1$ is much more practical than the case of $\alpha\geq1$ as
one usually allocates much less power to the reverse (feedback
pilot) channel than the forward transmission channel.

Case B) $\alpha \geq \frac{1}{M-1},\  N>1$: For such MIMO channel,
according to (\ref{eq14}), the additional diversity gain is $\Delta
_{d}(r)=(M^2N^2\alpha+r-r^2)$, for $r=0,1,...,N$. Specifically,
$\Delta _{d}(0)=M^2N^2\alpha$ and $\Delta _{d}(N)=\alpha
M^2N^2+N-N^2> MN^2+N$. If $0<r<N$, the additional diversity gain is
between the two extreme values $\Delta_{d}(0)$ and $\Delta_{d}(N)$.

Case C) $\alpha < \frac{1}{M-1},\  N>1$: When $r=N$,
$\Delta_{d}(N)=d(N)\geq d_1(N)=\alpha N(M-N+1)^2$. When $r<N$, for
the convenience of comparison with \cite{IEEErelay:DMT}, we consider
integer spatial multiplexing gains, i.e., $r=N-k,\ k=1,2,...,N$.
Since $r=N-k\leq (N-k){\tau}(k)$,  from Theorem 1, the achievable
diversity gain is $d(r)\geq k(M-N+k){\tau}(k)=
(M-r)(N-r)\left(1+\alpha (M-r)(N-r)\right)$. Recall that the optimal
diversity gain without CSIT
 is $d^*(r)=(M-r)(N-r)$. Therefore, the
additional achievable diversity gain with our scheme is
$\Delta_{d}(r)\geq \alpha (M-r)^2(N-r)^2=\alpha
\left(d^*(r)\right)^2$. It indicates that even a very small $\alpha$
leads to a significant diversity gain improvement.

We use numerical results to show the additional diversity gain
achieved with imperfect CSIT. We compare the following two
scenarios: 1) No CSIT \cite{IEEErelay:DMT};
 and 2)
Imperfect CSIT with power adaptation. Figs. 2 and 3 plot the DMT
curves for $3\times 3$ and $4\times 2$ MIMO fading channels,
respectively. It is obvious that imperfect CSIT provides significant
additional diversity gain improvement and offers non-zero diversity
gain at any possible spatial multiplexing gain. Fig.~2 also shows
the impact of $\alpha$ value. When
$\alpha\geq\frac{1}{M-1}=\frac{1}{2}$, we only have $d_N(r)<\infty$
and thus $\mathcal{B}=\{3\}$. Therefore, the DMT curve is a single
line segment. However, when $\alpha=\frac{1}{3}<\frac{1}{M-1}$,
$\mathcal{B}=\{1,2,3\}$. Therefore, the DMT curve is made up of
three discontinuous line segments. Fig.~3 shows how
$d_{\mathcal{O}}(r)$ depends on $d_1(r)$ and $d_2(r)$ in a $4\times
2$ MIMO channel with $\alpha=0.1$. We observe that $d_2(r)\geq
d_1(r)$ and there is only one corner point on $d_1(r)$ (or $d_2(r)$)
over spatial multiplexing gain region $r\in \tilde{\Omega}_1$ (or
$r\in \tilde{\Omega}_2$).

%

Next we illustrate the impact of $\alpha$ on DMT. Fig.~4
 plots the relationship between
the achievable diversity gain and the channel quality $\alpha$ in a
MISO/SIMO channel. It clearly shows that power adaptation makes
better use of the imperfect CSIT than rate adaptation. In other
words, to achieve the same performance our scheme saves a great
amount of pilot power and thus is more applicable. Specifically, the
diversity gain improvements over \cite{IEEErelay:DMT} and
\cite{IEEErelay:OFT} are $M^2\alpha$ and $(M-1)M\alpha$,
respectively, at any spatial multiplexing gain.
It is no doubt that the achievable DMT increases with CSIT quality
$\alpha$. Fig.~5 plots how the achievable diversity gain with power
adaptation improves with the channel quality $\alpha$ in a $5\times
3$ MIMO channel at the full multiplexing gain. We observe that there
are fast increases of diversity gain at $\alpha=0.25$ and
$\alpha=0.1667$. These two values of $\alpha$ are actually
thresholds for $d_k(r)<\infty, k=1,2,3$. When $\alpha\geq
\frac{1}{M-1}=0.25$, $\mathcal{B}=\{3\}$. Therefore, we have
$d_{\mathcal{O}}(N)=d_3(N)$. When $0.1667 \leq \alpha<0.25$, we have
$\mathcal{B}=\{2,3\}$ and $d_{\mathcal{O}}(N)=d_2(N)$. When
$\alpha<0.1667$, we have $\mathcal{B}=\{1,2,3\}$ and
$d_{\mathcal{O}}(r)=d_1(N)$. Combining with the fact that
$d_1(r)<d_2(r)<d_3(r)$ for any fixed $\alpha$, it is not difficult
to understand the cliffs on this curve.

Note that the additional diversity gain comes at the price of
reverse channel pilot power to obtain the CSIT. As long as the
reverse channel SNR does not scale with $\rho$, i.e., $\alpha=0$,
even with some partial CSIT at the transmitter, there will be no
improvement on the fundamental DMT. However,
 when the reverse channel SNR relative to $\rho$ becomes asymptotically zero, i.e., $\alpha<1$,
 there will be a significant improvement of the diversity gain.
   When the reverse channel SNR as compared to the forward SNR is asymptotically unbounded,
   i.e., $\alpha>1$, one can achieve the full spatial multiplexing gain while enjoying a
   even more remarkable diversity.

\section{Conclusion}

In this paper, we investigated the impact of CSIT on the tradeoff of
diversity and spatial multiplexing in MIMO fading channels. For
MISO/SIMO channels, we showed that using power adaptation, one can
achieve a diversity gain of $d(r)=K(1-r+K\alpha)$, where $K$ is the
number of transmit antennas in the MISO case or the number of
receive antennas in the SIMO case. This is not only higher but also
more efficient than the available results in literature. For general
MIMO channels  with $M>1$ transmit and $N>1$ receive antennas, when
$\alpha$ is above some certain threshold, one can achieve a
diversity gain of $d(r)=MN(1+MN\alpha)-(M+N-1)r$; otherwise, the
achievable DMT is a collection of discontinuous line segments
depending on $M$, $N$, $r$ and $\alpha$. The presented DMT shows
that exploiting imperfect CSIT through power adaptation
significantly increases the achievable diversity gain in MIMO
channels.

\appendices
\section{}
Letting $q_n\triangleq-\log(b_n)/\log (\rho)$ for all $n$ and
$\mathbf{q}\triangleq[q_1,...,q_N]$, we have
\begin{equation}
\begin{split}
&E\{P(\hat{\textbf{H}})\} =\int_{\textbf{b}\succeq
0}\frac{\kappa\bar{P} }{\left(\prod_{n=1}^N b_n^{2n-1+M-N}\right)^t}
\hat{\xi}^{-1}\prod_{n=1}^N b_n^{M-N} \prod_{n<j}^N
(b_n-b_j)^2\exp\left(-\frac{1}{1+\sigma_e^2}\sum_{n=1}^N b_n\right)
d \textbf{b}\\
&=\int_{\textbf{q}\succeq 0} \frac{\kappa\bar{P} \hat{\xi}^{-1}(\log
\rho)^N}{\left(\prod_{n=1}^N \rho^{-(2n-1+M-N)q_n}\right)^t}
 \prod_{n=1}^N \rho^{-(M-N+1)q_n} \prod_{n<j}^N (\rho^{-q_n}-\rho^{-q_j})^2\exp\left(-\frac{1}{1+\sigma_e^2}\sum_{n=1}^N \rho^{-q_n}\right)
d \textbf{q}.
\end{split}
\label{eq13}
\end{equation}


At high SNRs, it is easy to show that
\begin{equation}
E\{P(\hat{\textbf{H}})\}= \lim_{\rho\rightarrow\infty}
\int_{\textbf{q}\succeq 0}\kappa  \bar{P}  \hat{\xi}^{-1}(\log
\rho)^N \left(\prod_{n=1}^N \rho^{-(2n-1+M-N)q_n}\right)^{(1-t)} d
\textbf{q}=\bar{P}.
\end{equation}

\section{}
Let $v_{1,k_1},...,v_{k_1,k_1}$ denote the solutions of
$v_1,...,v_{k_1}$ that minimize $d_{k_1}(r)$, and let
$v_{1,k_2},...,v_{k_2,k_2}$ denote the solutions of
$v_1,...,v_{k_2}$ that minimize $d_{k_2}(r)$. Without loss of
generality, we assume
\begin{eqnarray}
&&v_{1,k_1}=...=v_{i-1,k_1}=\tau(k_1),\tau(k_1)>v_{i,k_1}\geq \alpha, v_{i+1,k_1}=...=v_{k_1,k_1}=\alpha\\
&&v_{1,k_2}=...=v_{j-1,k_2}=\tau(k_2),\tau(k_2)>v_{j,k_2}\geq \alpha, v_{j+1,k_2}=...=v_{k_2,k_2}= \alpha.
\end{eqnarray}

It follows that the corresponding spatial multiplexing gain $r$
satisfies
\begin{eqnarray}
 &&r=(N-i+1)\tau(k_1)-(k_1-i)\alpha-v_{i,k_1}\\ &&r=(N-j+1)\tau(k_2)-(k_2-j)\alpha-v_{j,k_2}
 \end{eqnarray}which leads to
\begin{equation}
\label{eqnv2v1}
\left\{(N-j+1)\tau(k_2)-(k_2-j)\alpha-v_{j,k_2}\right\}
-\left\{(N-i+1)\tau(k_1)-(k_1-i)\alpha-v_{i,k_1}\right\}=0.
\end{equation}

We consider the following three cases.\\
Case 1) $j<i$: Letting $B$ denote the LHS of (\ref{eqnv2v1}), we
have
\begin{equation}
\begin{split}
B>&
(N-j)\tau(k_2)-(k_2-j)\alpha  -(N-i+1)\tau(k_1)-(k_1-i+1)\alpha\\
\geq &(N-i+1)\tau(k_2)-(k_2-i+1)\alpha  -(N-i+1)\tau(k_1)-(k_1-i+1)\alpha\\
\geq &\alpha(k_2-k_1)\left((N-i+1)(M-N+k_2+k_1)-1\right)>0.
\end{split}
\end{equation}
This contradicts with $B=0$. Therefore, $j<i$ is not possible.

Case 2) $j>i$: It is easy to observe that
\begin{eqnarray}
&&v_{1,k_2}-v_{1,k_1}=...=v_{i-1,k_2}-v_{i-1,k_1}
=k_2\alpha(M-N+k_2)-k_1\alpha(M-N+k_1)>0\\
&&v_{i,k_2}-v_{i,k_1}>k_2\alpha(M-N+k_2)-k_1\alpha(M-N+k_1)>0\\
&&v_{i+1,k_2}-v_{i+1,k_1},...,v_{k_1,k_2}-v_{k_1,k_1}\geq \alpha-\alpha= 0\\
&&v_{k_1+1,k_2},...,v_{k_2,k_2}\geq \alpha.
\end{eqnarray}

Then, it follows that
\begin{equation}
d_{k_2}(r)-d_{k_1}(r)=\sum_{i=1}^{k_2}v_{i,k_2}-\sum_{i=1}^{k_1}v_{i,k_1}>0.
\end{equation}

Case 3) $j=i$: Similarly, we have
\begin{eqnarray}
\label{eqn_65}
&&v_{1,k_2}-v_{1,k_1}=...=v_{i-1,k_2}-v_{i-1,k_1}=k_2\alpha(M-N+k_2)-k_1\alpha(M-N+k_1)>0\\
\label{eqn_66}
&&v_{i+1,k_2}=...=v_{k_2,k_2}=v_{i+1,k_1}=...=v_{k-1,k_1}=\alpha.
\end{eqnarray}

From (\ref{eqnv2v1}), we get
\begin{equation}
\label{eqn_67}
v_{i,k_2}-v_{i,k_1}=\alpha(k_2-k_1)\left((N-i+1)(M-N+k_2+k_1)-1\right)>0.
\end{equation}
Combining (\ref{eqn_65}), (\ref{eqn_66}) and (\ref{eqn_67}), we get
$d_{k_2}(r)>d_{k_1}(r)$. The proof of Lemma 2 is complete.

\section{}
We compare the spatial multiplexing gain $r$ of the corner point $k'$ ($k'=1,...,k-1$) on the DMT curve $d_k(r)$, i.e., $r=(N-k')\tau(k)-(k-k')\alpha$, with the lower boundary of
$\Omega_{k-1}$, i.e., $(N-k+1)\tau(k-1)$, and get
\begin{equation}
(N-k')\tau(k)-(k-k')\alpha - (N-k+1)\tau(k-1) \geq
\left((N-k+1)(M-N+2k-1)-1\right)\alpha > 0.
\end{equation}

If $(N-k')\tau(k)-(k-k')\alpha<N$, it suffices to have
$(N-k')\tau(k)-(k-k')\alpha\in {\Omega}_{k-1}$. Otherwise, we get
$(N-k')\tau(k)-(k-k')\alpha \notin \Omega_k$. Since
${\Omega}_{k-1}\bigcap \tilde{\Omega}_{k}=\emptyset$ and
$\tilde{\Omega}_{k}\subseteq \Omega_k$, both cases lead to
$r\notin\tilde{\Omega}_k$. This completes the proof of Lemma 3.

\newpage

\begin{figure}[!t]
\centering
  \includegraphics[width=3.5in]{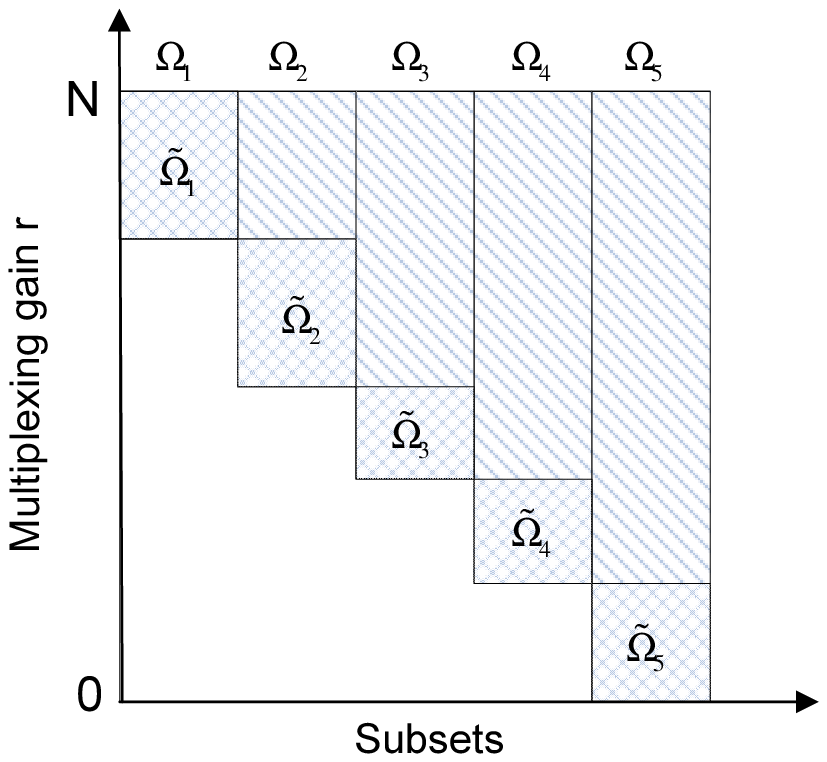}
  \caption{Relationship of $\Omega_k$ and $\tilde{\Omega}_k$.}
  \label{fig_1} 
\end{figure}


\begin{figure}[!t]
\centering
  \includegraphics[width=5in]{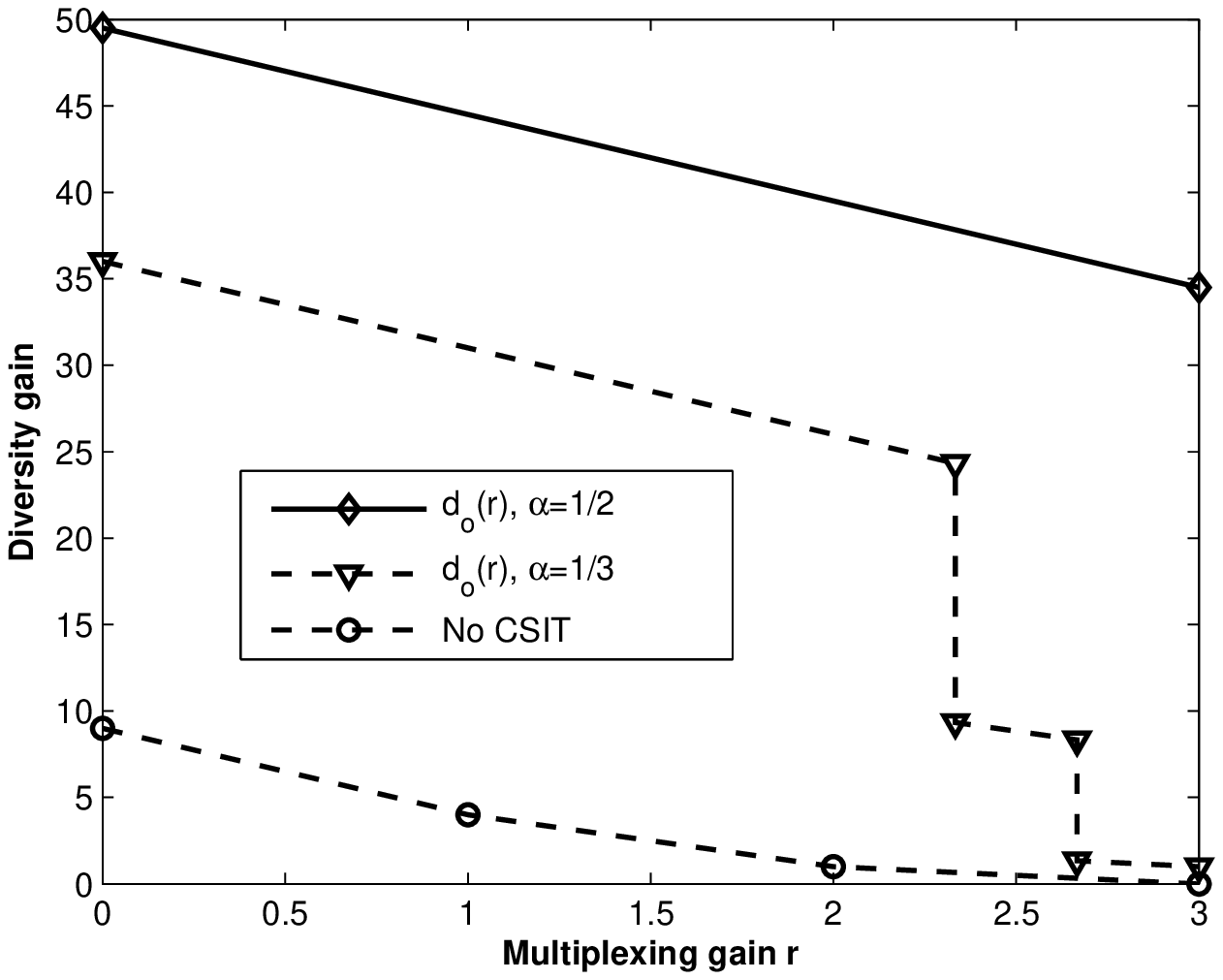}
  \caption{DMT in a $3 \times 3$ MIMO channel. Note that $\mathsf{d_o(r)}$ in the legend denotes $d_{\mathcal{O}}(r)$.}
  \label{fig_3} 
\end{figure}

\begin{figure}[!t]
\centering
  \includegraphics[width=5in]{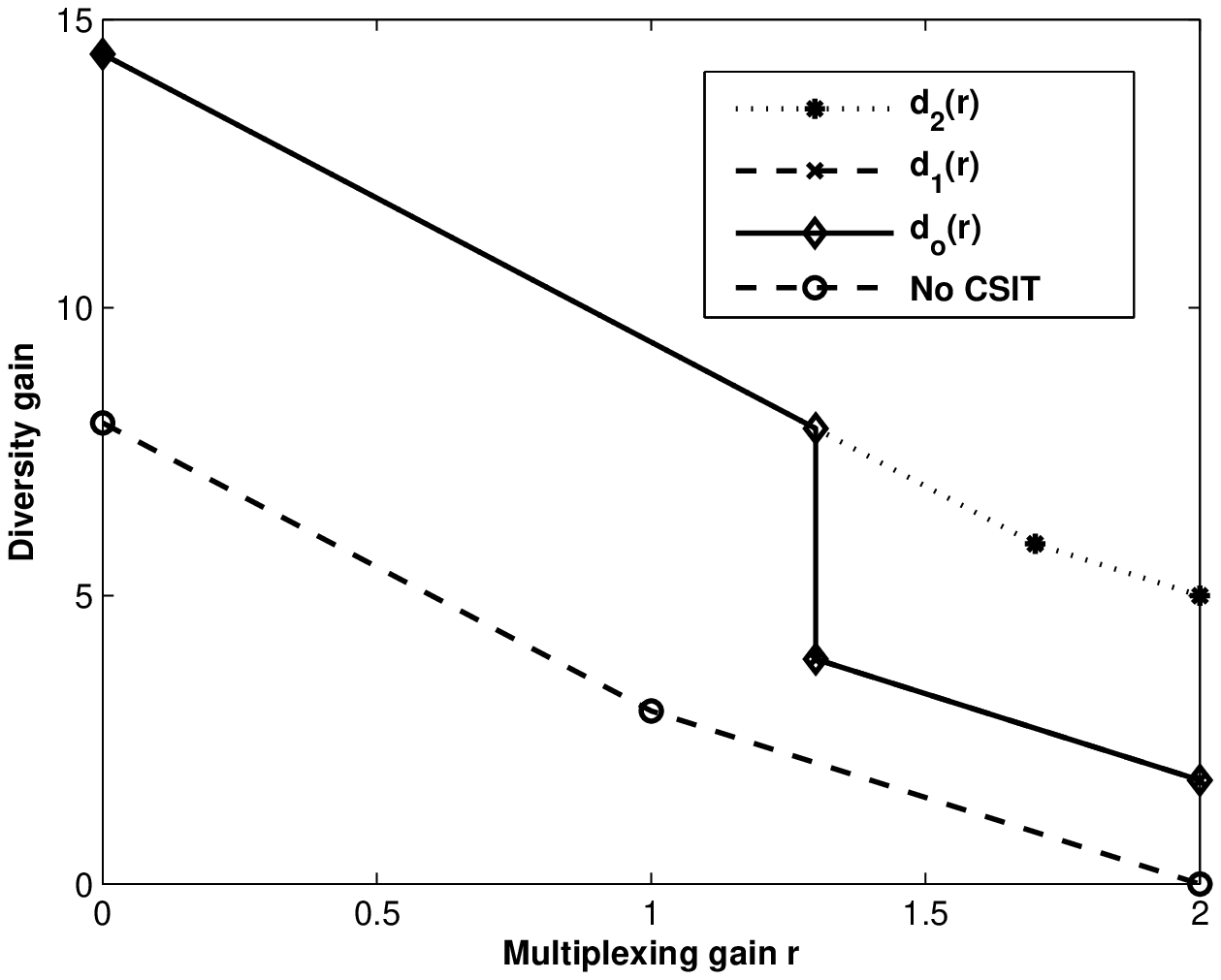}
  \caption{DMT in a $4\times 2$ MIMO channel with $\alpha=0.1$. Note that $\mathsf{d_o(r)}$ in the legend denotes $d_{\mathcal{O}}(r)$.}
  \label{fig_4} 
\end{figure}

\begin{figure}[!t]
\centering
  \includegraphics[width=5in]{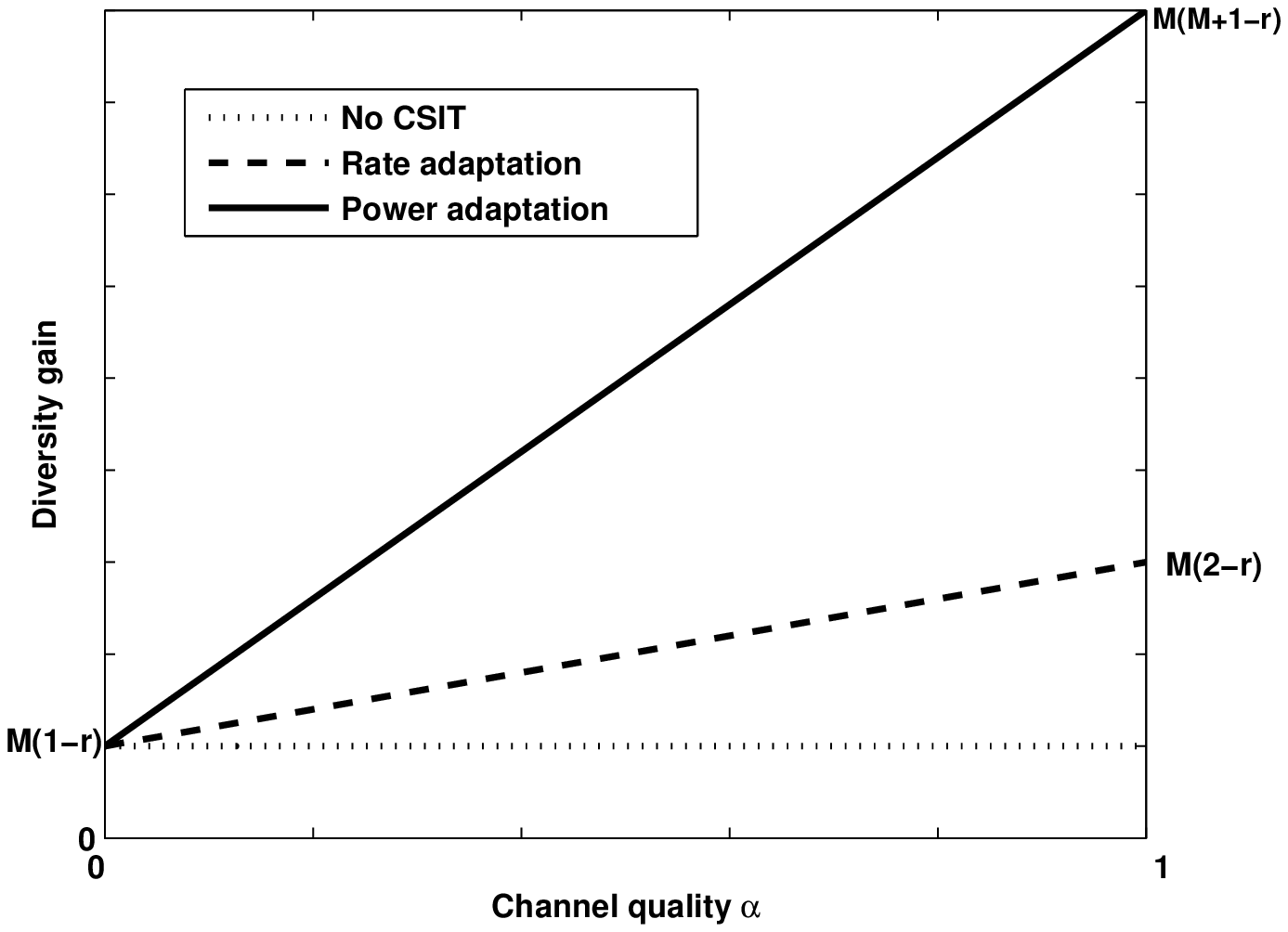}
  \caption{Diversity gain versus channel quality $\alpha$ in a SIMO/MISO channel.}
  \label{fig_5} 
\end{figure}

\begin{figure}[!t]
\centering
  \includegraphics[width=5in]{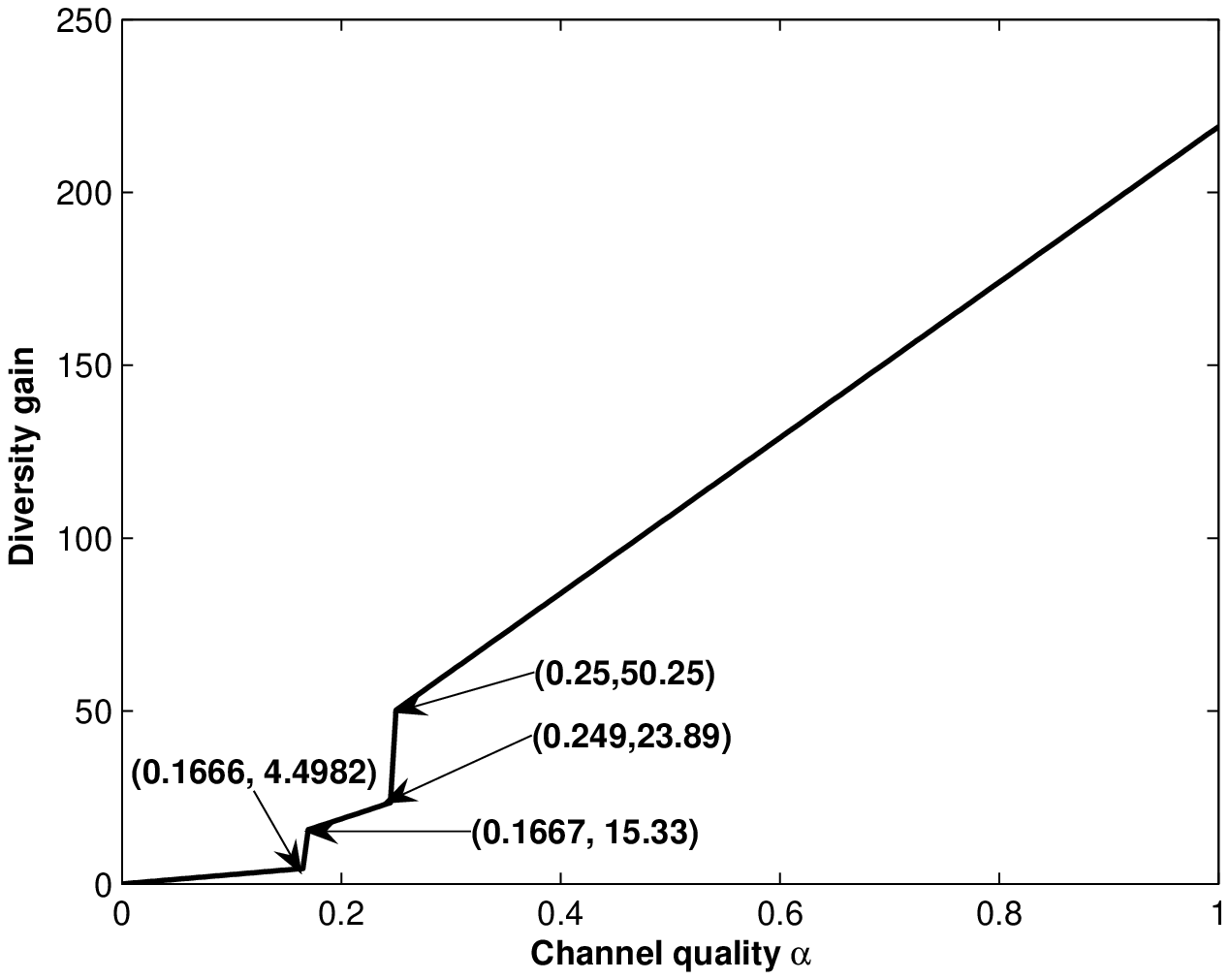}
  \caption{Diversity gain at $r=N$ versus channel quality $\alpha$ in a $5 \times 3$ MIMO channel.}
  \label{fig_6} 
\end{figure}

\end{document}